\def\year{2018}\relax

\documentclass[letterpaper]{article} 
\usepackage{aaai19}  
\usepackage{times}  
\usepackage{helvet}  
\usepackage{courier}  
\usepackage{graphicx}  
\usepackage{booktabs} 
\usepackage{multirow}
\usepackage{amsmath}
\usepackage{amsfonts}
\usepackage{enumitem}
\usepackage{xcolor}
\usepackage{hyperref}

\usepackage{footmisc}
\usepackage{todonotes}
\usepackage{subcaption}
\usepackage{colortbl}

\title{Spread of hate speech in online social media}
\author{Binny Mathew \quad Ritam Dutt \quad Pawan Goyal \quad Animesh Mukherjee\\
	Indian Institute of Technology, Kharagpur \\
	{\tt \small binnymathew@iitkgp.ac.in, ritam.dutt@gmail.com} \\
	{\tt \small pawang.iitk@gmail.com , animeshm@gmail.com}
}

\begin{document}
	
\maketitle

\begin{abstract}
The present online social media platform is afflicted with several issues, with hate speech being on the predominant forefront. The prevalence of online hate speech has fuelled horrific real-world hate-crime such as the mass-genocide of Rohingya Muslims, communal violence in Colombo and the recent massacre in the Pittsburgh synagogue. Consequently, It is imperative to understand the diffusion of such hateful content in an online setting. We conduct the first study that analyses the flow and dynamics of posts generated by hateful and non-hateful users on Gab (gab.com) over a massive dataset of 341K users and 21M posts. Our observations confirms that hateful content diffuse farther, wider and faster and have a greater outreach than those of non-hateful users. A deeper inspection into the profiles and network of hateful and non-hateful users reveals that the former are more influential, popular and cohesive. Thus, our research explores the interesting facets of diffusion dynamics of hateful users and broadens our understanding of hate speech in the online world. 
\end{abstract}


\section{Introduction}

The Internet is one of the greatest innovations of mankind which has brought together people from every race, religion, and nationality. Social media sites such as Twitter and Facebook have connected billions of people\footnote{\url{https://techcrunch.com/2018/07/25/facebook-2-5-billion-people}} and allowed them to share their ideas and opinions instantly. That being said, there are several ill consequences as well such as online harassment, trolling, cyber-bullying, and hatespeech. 

Twitter defines hatespeech\footnote{\label{twitter_hate_policy}\url{https://help.twitter.com/en/rules-and-policies/hateful-conduct-policy}} as any tweet that `promotes violence against other people on the basis of race, ethnicity, national origin, sexual orientation, gender, gender identity, religious affiliation, age, disability, or serious disease. Even though several government and social media sites are trying to curb the hatespeech, it is still plaguing our society. Facebook has been blamed by United Nations investigators in playing a leading role in the possible genocide of the Rohingya community in Myanmar by spreading hatespeech\footnote{\url{https://www.reuters.com/investigates/special-report/myanmar-facebook-hate}}. Sri Lanka has also accused Facebook for instigating anti-Muslim mob violence that left three people dead\footnote{\url{https://goo.gl/QMU8e7}}. With hate crimes increasing in several states\footnote{\url{https://goo.gl/9gAjDg}}, there is an urgent need to have a better understanding of how hateful posts spreads in online social media.

In this paper, we perform the first study which looks into the diffusion dynamics of hate in online social media. We choose Gab(Gab.com) for all our analysis. This choice is primarily motivated by the nature of Gab. Unlike other social media sites such as Twitter and Facebook, Gab promotes ``free speech'' and allows users to post contents that may be hateful in nature without any fear of repercussion. This has led to the migration of several Twitter users who were banned/suspended for violating its terms of service, namely for abusive and/or hateful behavior~\cite{zannettou2018gab}. This provides a unique opportunity to study how the hateful content would spread in the online medium, if there were no restrictions. 

To this end, we crawl the Gab platform and acquire 21M posts by 341K users over a period of 20 Months (October, 2016 to June, 2018). Our analysis reveals that the posts by hateful users tend to spread faster, farther, and wider as compared to normal users. Our main contributions are as follows-

\begin{itemize}
	\item We perform the first study which looks into the diffusion dynamics of posts by hateful accounts.

	\item We find that the hate users in our dataset (which constitutes 0.3\% of the total number of users) are very densely connected and are responsible for 18.65\% of posts generated in Gab.

\end{itemize}

In summary, our analysis reveals that the hatespeech has a much higher spreading velocity. The posts of hateful users receive much larger audience and as well at a faster rate. As a case study, we also investigate the detailed account characteristics of Robert Gregory Bowers, the sole suspect of the Pittsburgh synagogue shooting\footnote{\label{pittsburg_shooting}\url{https://en.wikipedia.org/wiki/Pittsburgh_synagogue_shooting}}.

\section{Dataset Description}

In order to understand the diffusion dynamics in Gab, we collect a massive dataset of posts and users by following the crawling methodology in ~\cite{zannettou2018gab}. We use Gab's API to crawl the site in a snowball methodology. We first obtain the data for the most popular user as returned by Gab's API and then collect the data for all their followers and followings. We collect the following types of information: 1) basic details about each user like username, score, account creation date; 2) all the posts of each user; 
3) all the followers and followings for each users. This resulted in a massive dataset whose details are presented in Table~\ref{tab: dataset-details}. We have only collected the publicly available data posted in Gab and make no attempt to de-anonymize the users. We outline the procedure to distinguish between hateful and non-hateful users in the following section.

\begin{table}[htb]
\centering
\begin{tabular}{l l}
\hline
Property & Value\\ \hline \hline
Number of posts&21,207,961\\
Number of reply posts&6,601,521\\
Number of quote posts&2,085,828\\
Number of reposts&5,850,331\\
Number of posts with attachments &9,669,374\\
Number of user accounts&341,332\\
Average follower per account & 62.56 \\
Average following per account & 60.93\\
\hline
\end{tabular}
\caption{Description of the dataset.}
\label{tab: dataset-details}
\end{table}

\subsection{Identifying hateful content}

Gab has been at the center of several hate activity. With the recent Pittsburg shooting, and removal of the app from play store, it has become quite infamous.  The volume of hateful content is Gab is 2.4 times higher than that of Twitter~\cite{zannettou2018gab} which justifies our choice of Gab. We adopted a multi step approach to curate our dateset.

\noindent\textbf{Lexicon based filtering}:
We created a lexicon\footnote{The lexicon is available here: \url{https://goo.gl/8iHTDP}} of 45 high-precision unigrams and bigrams that are often associated with hate like `kike' (slur against Jews), `paki' (slur against Muslims), `beached whale' (slur against fat people). These hate words were initially selected from the Hatebase\footnote{\label{hatebase}\url{https://www.hatebase.org}} and Urban dictionary\footnote{\url{https://www.urbandictionary.com}}. Words such as `banana', `bubble' are present in hatebase which could easily appear in benign context. In order to avoid ambiguity, we ran multiple iterations and carefully chose those keywords which were not ambiguous in Gab. 

We leverage these high precision keywords to identify explicit hate posts based on their textual content. The total number of unique posts which have been identified explicitly as 'Hate' were 167,782 or 0.79\% of the entire dataset. However, since posts need not necessarily contain solely textual information (45.59\% of all posts include an attachment in the form of images, videos, and URLs), we resort to a diffusion based model of identifying hate users in the social network. 

\noindent\textbf{Identifying hateful users}:
Using the high precision lexicon would miss out on several users who might be hateful in nature but are not selected as they did not post any content with words from our lexicon (like using images and videos). In order to capture such obscure hate users, we leverage the methodology used by~\cite{riberio18}.
We enumerate the steps of our methodology below.

\begin{itemize}
\item We identify the initial set of hateful users as those who have written at least 10 posts, with at least one hateful keyword in each of them. This results in a set of 1863 hateful users.
\item We create a repost network where nodes represents the users and edge-weights denotes posting and reposting frequency. We convert the repost network into a belief network by reversing the edges in the original network and normalizing the edge weights between 0 and 1. We explain this further in the subsequent section.
\item We then run a diffusion process based on the DeGroot's learning model~\cite{golub2010naive} on the belief network. We assign an initial belief value of 1 to the 1863 users identified earlier and 0 to all the other users. The diffusion model aims to identify users who did not explicitly use any of the hateful keywords, yet have a high potential of being a hate user due to homophily.
\item We observe the belief values of all the users in the network after \textit{five} iterations of the diffusion process and divide the users into four strata, $[0, .25)$, $[.25, .50)$,
$[.50, .75)$ and $[.75, 1)$ according to their associated belief. 
\end{itemize}

We define users whose belief values lie within $[.75,1]$ as hateful and those whose belief values lie within $[0,.25)$ as non-hateful with the additional constraint that each of these users should have at least \textit{five} posts. We do so since it is difficult to judge a person on the basis of a single post. We thus obtain a set of 1055 hateful users and 62827 non-hateful users, which comprises 0.3\% and 18.406\% of the entire dataset. We refer to the set of hateful and non-hateful users as KH (read 'Known hateful user') and NH (read 'Not hateful user') respectively henceforth.

\begin{figure}[h]   	
	\centering
	\begin{subfigure}[b]{0.15\textwidth}
		\includegraphics[width=\textwidth]{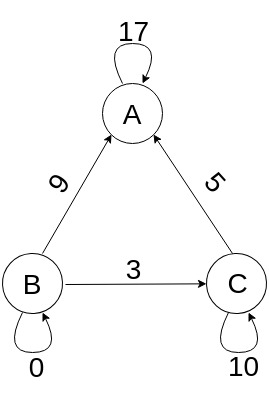}
		\caption{\label{fig:repost-network}Repost Network}
		
	\end{subfigure}
	\begin{subfigure}[b]{0.15\textwidth}
		\includegraphics[width=\textwidth]{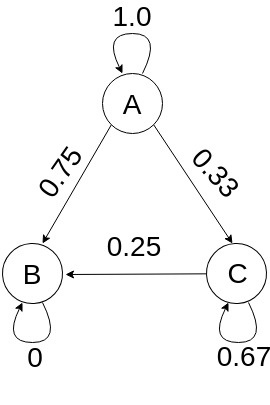}	
		\caption{\label{fig:belief-network}Belief Network}	
	\end{subfigure}
    \begin{subfigure}[b]{0.15\textwidth}
		\includegraphics[width=\textwidth]{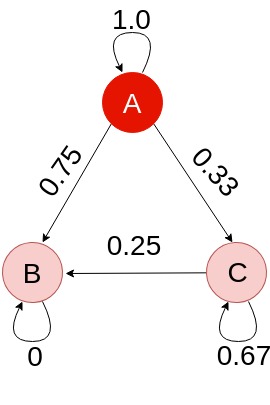}
		\caption{\label{fig:belief-diffusion}Belief diffusion}	
	\end{subfigure}
    \caption{\label{fig:degroot-diffusion}Description of the DeGroot's model for information diffusion in a toy network.}
\end{figure}

\noindent\textbf{DeGroot's model of information diffusion}: We illustrate a repost network with three users (A, B, C) in Figure~\ref{fig:repost-network}. An edge-weight of 9 from B to A denotes that user B has reposted 9 posts of A while a self loop of A of weight 17 denotes that A has posted 17 times. We convert the repost network into a diffusion network as shown in~\ref{fig:belief-network} by reversing the edges, with the edge-weights normalized. The edge weights are normalized by dividing the edge weight from C to A in the original network by the sum of the edge weights originating from C (including self loops). For example, user C in Figure~\ref{fig:repost-network} has reposted A 5 times and has posted 10 times. Thus the value of edge weights from A to C is $\frac{5}{15}$ or 0.33 and the weight of the self-loop at C is $\frac{10}{15}$ or 0.67 as shown in Figure~\ref{fig:belief-network}. The normalized edge-weight is a measure of the user's belief being influenced by her neighbors. Let us denote the belief of A, B and C at the time instant $i$ as $b_{A}^{i}, b_{B}^{i}, b_{C}^{i}$ respectively. The belief of user C at time instant $i+1$ can be written as 
\begin{equation}
b_{C}^{i+1}= 0.33\times b_{A}^{i}+ 0.67\times b_{C}^{i}
\end{equation}
Thus belief propagation takes place in an iterative fashion using the DeGroot's model. If we consider the initial beliefs of A, B and C to be 1, 0 and 0 respectively, their corresponding beliefs at time instant 1 would be 1, 0.75 and 0.33 as demonstrated in Figure~\ref{fig:belief-diffusion}.

\subsection{Dataset evaluation}

We evaluate the quality of the final dataset of hateful and non-hateful accounts through human judgment. We ask four annotators to determine if a given account is hateful or non-hateful as per their perception. The annotators consisted of three undergraduate students with major in Computer Science and one PhD student in Social Computing. Since Gab does not have any policy for hatespeech, we use the guidelines defined by Twitter~\footref{twitter_hate_policy} for this task. We provide the annotators with a class balanced random sample of 200 user accounts~\footnote{We have used a random sample of 200 accounts per class to keep the monetary cost manageable}. Each account was evaluated by two independent annotators.

We observe that the two annotators found 86.9\% and 93.2\% of the hate accounts from our sample as hateful, yielding a substantial high Cohen's $\kappa$ score of 0.69. Likewise 92.2\% and 99.4\% of the non-hateful accounts from our sample were adjudged to be non-hateful yielding a very high $\kappa$ score of .87. These results show that the dataset generated by our method is of high quality with minimal noise.

\section{Diffusion dynamics of posts}
In this section, we observe the diffusion of information throughout the network and analyze the differences in diffusion of posts generated by the hateful users and those generated by the non-hateful users. 

\subsection{Model description}
\begin{figure}[h]   	
	\centering
    
    \hspace{-15mm}
	\begin{subfigure}[b]{0.20\textwidth}
 \includegraphics[width=\textwidth]{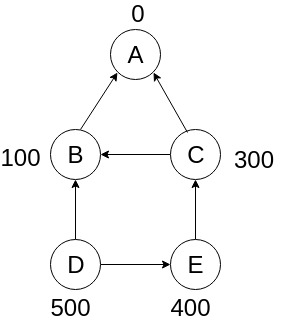}
		\caption{Repost graph}
		\label{fig:network}
	\end{subfigure}
	\begin{subfigure}[b]{0.20\textwidth}	\includegraphics[width=\textwidth]{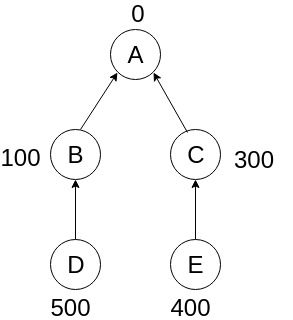}
		\caption{LRIF model}
		\label{fig:lrif}
	\end{subfigure}
\hspace{-18mm}    
     
    \caption{DAG generated by the LRIF model on a sample re-post network. The numbers indicate the time in seconds of reposting. The links are formed between User C and A since A posted earlier than B.}
    \label{fig:mr_lr_if models}
\end{figure}

We refer to the path traced by a post as it is reposted by other users as a cascade and the original user as the root user. Since it is not possible to trace the exact influence path, i.e., the user who influenced the reposting, we leverage the social network connections (followers and friends) as means of information diffusion and influence similar to~\cite{LRIF-2014}. In all the models, an edge is formed between two users if there exists a follower-following relationship between the users.
We deploy the Least Recent Influencer Model (LRIF)~\cite{bakshy2011everyone} to observe the information diffusion. Previous research~\cite{LRIF-2014,alrajebah2017deconstructing} have also used such models to study the diffusion of information in online social media. 
In the LRIF model, users are influenced by the first exposure to a message even if they do not act immediately. Essentially, the model seek to avoid exhaustive search by converting the network into a directed acyclic graph, thereby, reducing the time complexity. We illustrate the DAG generated by the LRIF models in Figure~\ref{fig:mr_lr_if models}. 
	The sample network is shown in Figure \ref{fig:network} comprising 5 users. A directed edge between any 2 users (say from B to A) specifies the follower-following relationship (B follows A). The number beside each user specifies the time of reposting, with A being the root user. The DAG generated by the LRIF model is shown in \ref{fig:lrif}.
    
\begin{table*}[!ht]
\centering
\resizebox{1.0\textwidth}{!}{%
\begin{tabular}{|c|c|c|c|c|c|c|c|c|c|}
\hline
 &\multicolumn{2}{c|}{Posts} &\multicolumn{2}{c|}{Attachments} &\multicolumn{2}{c|}{Posts in groups} &\multicolumn{2}{c|}{Posts in topics} \\ \hline
Feature& Mean KH& Mean NH & Mean KH& Mean NH& Mean KH& Mean NH & Mean KH& Mean NH\\ \hline
Size&1.27&1.22&1.33&1.25&1.43&1.41&1.60&1.55\\ \hline
Depth&0.13&0.09& 0.16&0.11& 0.23&	0.23& 0.26&	0.24\\ \hline
Breadth&1.12&1.11& 1.14&1.12& 1.17&1.15&1.25&1.27\\ \hline
Average Depth&0.12&0.08 & 0.15&0.10& 0.20&0.21& 0.22&0.21\\ \hline
Structural virality &0.14&0.10& 0.17&0.11& 0.23&0.23&	0.26&0.24 \\ \hline
\end{tabular}}
\caption{Diffusion characteristics of posts of the hateful and the non-hateful users.}
\label{tab: diffusion-characteristics-all}
\end{table*}

\subsection{Characteristic cascade parameters}
In order to characterize the cascades generated by KH and NH users, we employ the following features as used in~\cite{mit-media-2018}.
\begin{itemize}
\item\textbf{Size} represents the number of nodes in the DAG which are reachable from the root user. It corresponds to the total number of unique users involved in the cascade of the post.


\item\textbf{Depth} is the length of the largest path from the root node of the cascade. The depth of a cascade, $D$, with $n$ nodes is defined as
\begin{equation}
D = \max{(d_i)}, 0\leq i\leq n
\end{equation}
where $d_i$ is the depth of node $i$.

\item\textbf{Average depth} is the average path length of all nodes reachable from the root user. For a cascade with $n$ nodes, we define its average depth $(AD)$ as
\begin{equation}
AD = \frac{1}{n}\sum_{i=1}^{n} d_i
\end{equation}
where $d_i$ is the depth of the node $i$.

\item \textbf{Breadth} is the maximum no. of nodes present at any particular depth in the DAG.
\begin{equation}
B = \max{(b_i)}, 0\leq i \leq d
\end{equation}
where $b_i$ denotes the breadth of the cascade at depth $i$ and $d$ denotes the maximum depth of the cascade.

\item \textbf{Structural virality} as defined by~\cite{goel2015structural}, is the average distance between all pairs of nodes in the DAG, assuming the DAG to be a tree. It is simply the Weiner index.
\begin{equation}
SV = \frac{1}{n(n-1)} \sum_{i=1}^{n}\sum_{j=1}^{n}d_{ij}
\end{equation}
where $d_{ij}$ represents the length of the shortest path between nodes $i$ and $j$.
\end{itemize}

\subsection{Experiments on varied nature of posts }
All subsequent evaluation is carried out on the DAG generated by the LRIF model. We include only the posts of KH and NH users which are not classified as `quotes' or `replies' since such posts have a low rate of repost. We also observe the diffusion characteristics for posts having attachments (images or media content) separately since such posts are hypothesized to be more viral. The supposed virality is attributed to the appeal of an image/ meme over plain textual information. Finally, in order to observe the community perspective, we look into posts which have been posted in groups and topics. We report the mean score of the cascade features for the different experiments in Table~\ref{tab: diffusion-characteristics-all}. 

\begin{figure*}[htb]   	
	\centering
	\begin{subfigure}[b]{0.19\textwidth}
		\includegraphics[width=\textwidth]{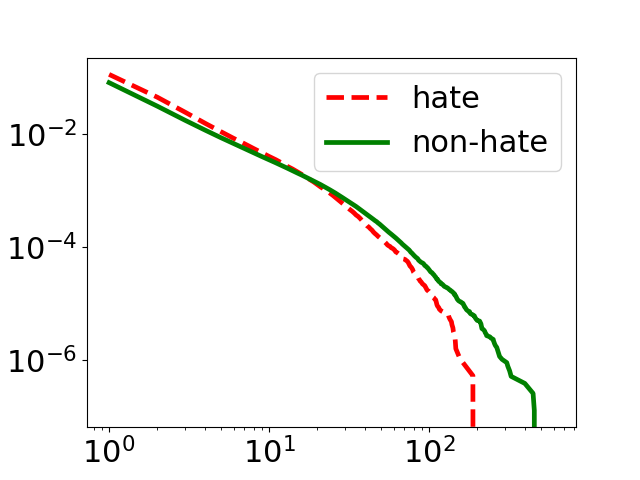}
		\caption{CCDF of size}
		\label{fig:size}
	\end{subfigure}
	\begin{subfigure}[b]{0.19\textwidth}
		\includegraphics[width=\textwidth]{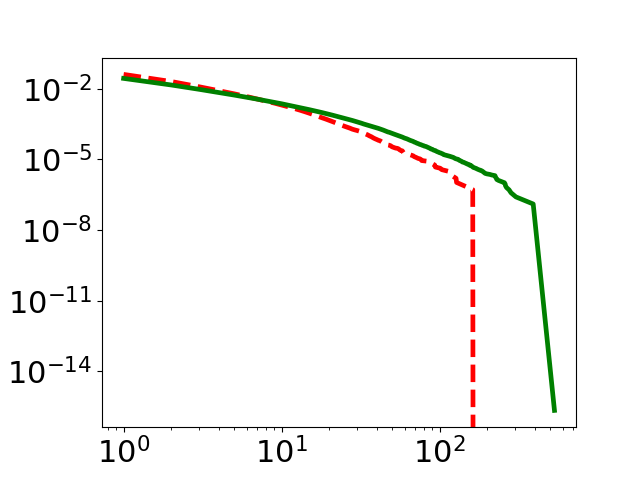}
		\caption{CCDF of breadth}
		\label{fig:breadth}
	\end{subfigure}
	\begin{subfigure}[b]{0.19\textwidth}
		\includegraphics[width=\textwidth]{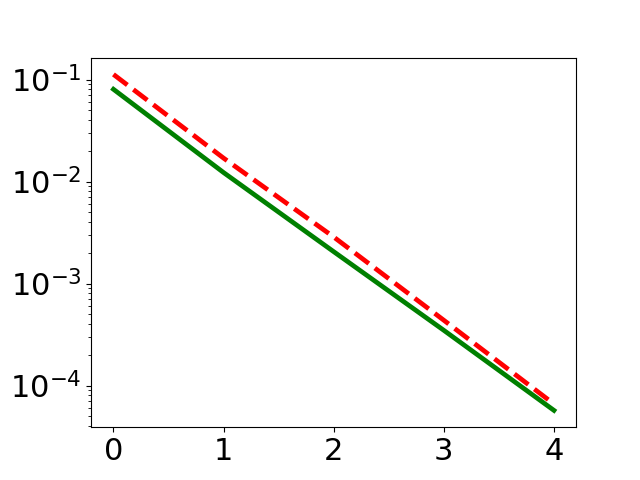}
		\caption{CCDF of depth}
		\label{fig:depth}
	\end{subfigure}
	\begin{subfigure}[b]{0.19\textwidth}
		\includegraphics[width=\textwidth]{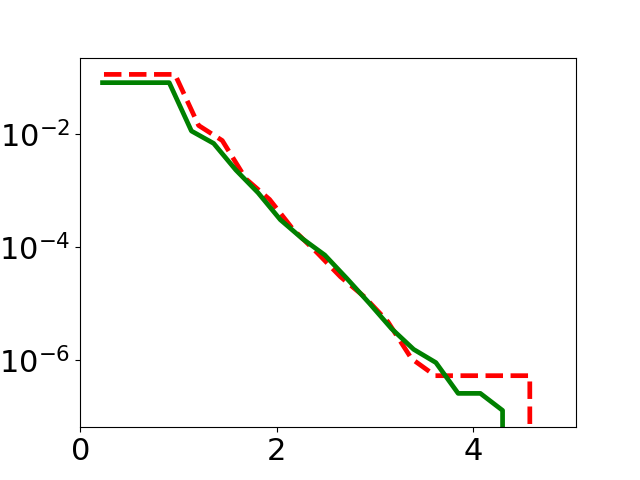}
		\caption{CCDF of average depth}
		\label{fig:avgdepth}
	\end{subfigure}
	\begin{subfigure}[b]{0.19\textwidth}
		\includegraphics[width=\textwidth]{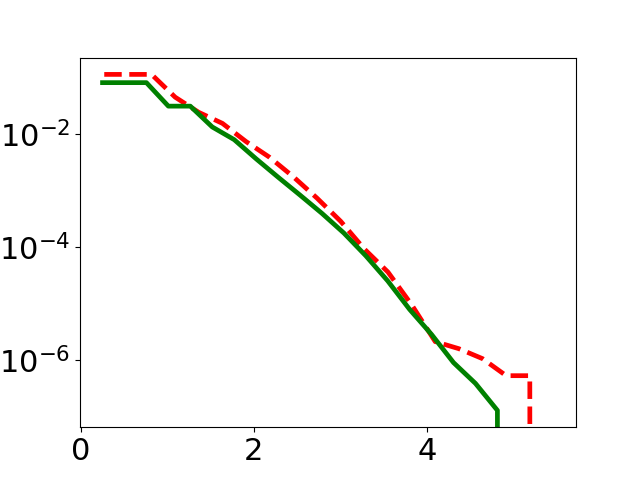}
		\caption{CCDF of virality}
		\label{fig: virality}
    \end{subfigure}
    
\caption{Different diffusion dynamics of the posts by hate and non-hate users using LRIF model. The cascade properties namely size, breadth, depth, average depth and virality are larger for the posts of hateful users.}
 \label{fig:diffusion-models}
\end{figure*}

\subsection{Characteristic differences in cascades of the hateful and non-hateful users}

\noindent\textbf{General cascade parameters}: The mean size (number of unique users) of a cascade is larger for posts of hateful users than non-hateful users. The CCDF of a cascade's size for both KH and NH users is shown in Figure~\ref{fig:size}. We observe that although the maximum size of a NH's cascade is larger, the cascade's size is significantly larger for KH users especially for the initial stages. Thus, the posts of hateful users have a larger audience. 

The mean breadth of a cascade is also larger for posts generated by hateful users implying that such posts spread wider (farther amongst a user's followers) than those generated by NH users. The CCDF of a cascade's breadth~\ref{fig:breadth} exhibits similar characteristics as a cascade's size. 

The mean depth, mean average depth and mean structural virality of a cascade are also significantly larger for posts generated by hateful users. Not only does it imply that such posts diffuse deeper into the network but they are also more viral~\cite{goel2015structural}. Moreover, as the CCDF for depth, avg-depth and virality (Figures~\ref{fig:depth},~\ref{fig:avgdepth} and~\ref{fig: virality} respectively) depicts, these properties remain consistently larger for the KH users throughout their entire distribution.

\noindent\textbf{Attachments and communal perspective}: It is observed that posts with attachments have a larger mean size, breadth, depth, average depth and structural virality implying that such posts have a greater outreach, diffuse wider, deeper and are more viral. This agrees with our hypothesis that attachments with memes and images are more instrumental in information diffusion than textual content. 

These diffusion properties becomes more pronounced when we consider only the cascades arising from posts in groups and topics. Groups and topics both represent sub-communities in GAB catered to a certain cause or serving a niche interest. The primary difference between groups and topics is that groups require prior approval for membership, implying that they are more exclusive. Hence, the number of users in a topic is more than those for a group which is reflected in the larger characteristic values of the cascade, as evident from Table~\ref{tab: diffusion-characteristics-all}.

The different characteristics manifest as significant ($p-value < 0.01$) according to the KS-test for posts, attachments and posts having a valid topic title. However, there is no significant difference between the diffusion characteristics of posts of hateful and non-hateful users within a group itself. This can be attributed to the exclusive nature of the group which ensures that users of similar nature are members thereby diminishing the individualistic differences in the posts. Likewise, due to their exclusivity the fractions of groups which abound in Gab is relatively small. The number of posts posted in groups is 1103K while those posted in topics are 976K.

\begin{figure}[tb]   	
	\centering
    \begin{subfigure}[b]{0.22\textwidth}
\includegraphics[width=\textwidth]{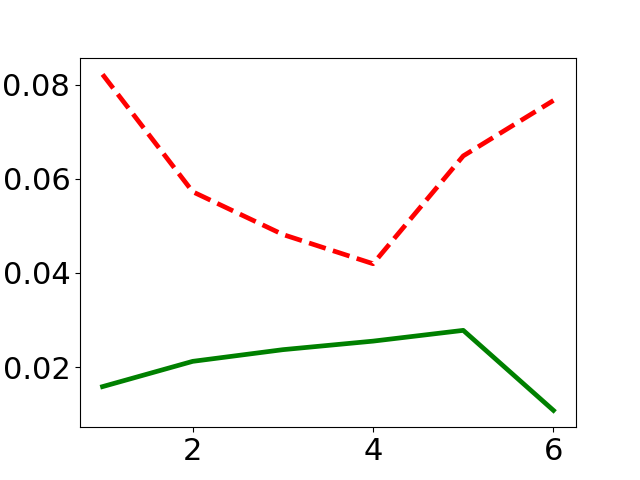}
		\caption{\% of KH propagators}
		\label{fig:depth-hate}
	\end{subfigure}
	\begin{subfigure}[b]{0.22\textwidth}	\includegraphics[width=\textwidth]{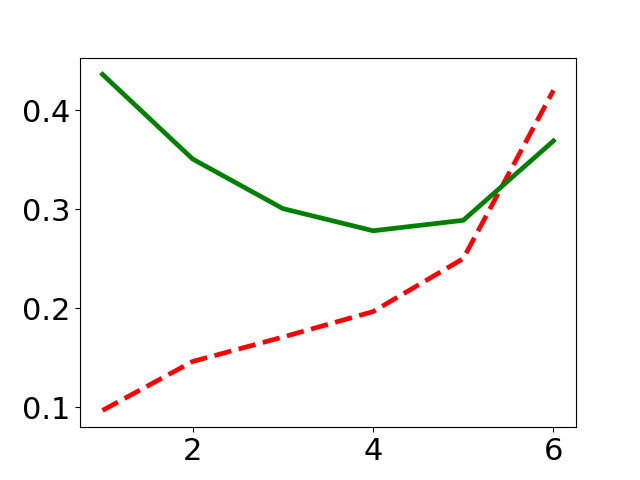}
		\caption{\% of NH propagators}
		\label{fig: depth-non-hate}
    \end{subfigure}
\caption{The proportion of hateful and non-hateful users who have reposted the root user across different depths. Hate users are early propagators for the posts of hateful users while non-hateful users are the early propagators for the posts of non-hateful users.}
 \label{fig:early-propagators}
\end{figure}

\noindent\textbf{Early adopters in a cascade}: Figures~\ref{fig:depth-hate} and~\ref{fig: depth-non-hate} illustrate the proportion of hateful and non-hateful propagators at each depth. It is evident that the hateful users are early adopters in the cascades of KH users, exhibiting strong degree of homophily. The reverse also holds true for non-hateful users who are the early adopters in the cascades of NH users. The change in monotonicity of the curves in both the diagrams after depth 4 can be attributed to the small number of cascades whose depth exceeded 4 levels (0.0065\% and 0.0057\% of KH and NH respectively). These are fast cascades where the information was propagated by a larger fraction of hateful users in KH posts and larger fraction of non-hateful users in NH posts.

\begin{figure*}[htb]   	
	\centering
    \hspace{-2mm}
	\begin{subfigure}[b]{0.19\textwidth}
		\includegraphics[width=\textwidth]{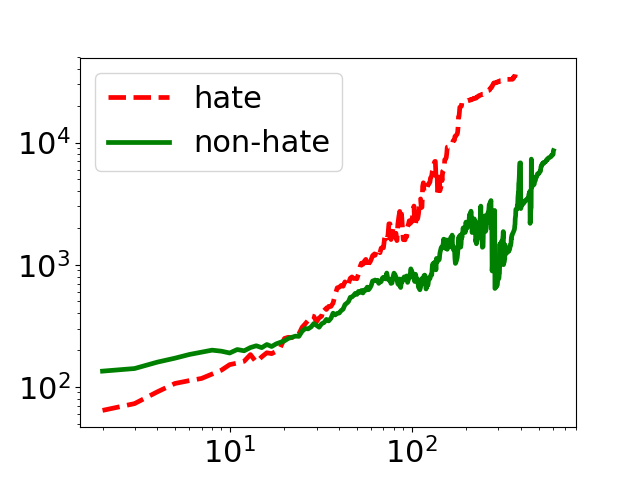}
		\caption{Size vs time}
		\label{fig:temp-size}
	\end{subfigure}
	\begin{subfigure}[b]{0.19\textwidth}
		\includegraphics[width=\textwidth]{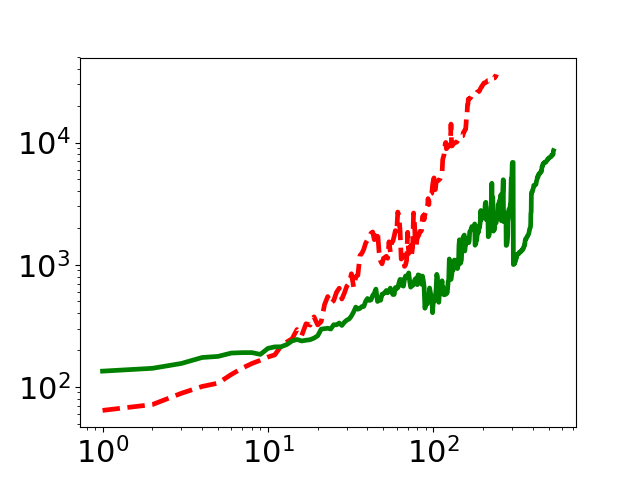}
		\caption{Breadth vs time}
		\label{fig:temp-breadth}
	\end{subfigure}
	\begin{subfigure}[b]{0.19\textwidth}
		\includegraphics[width=\textwidth]{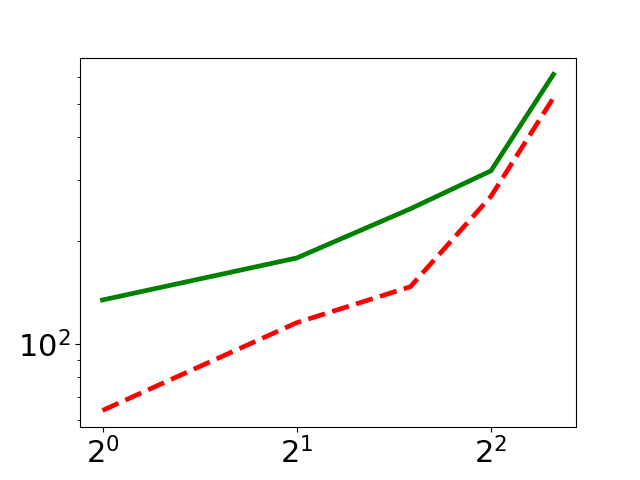}
		\caption{Depth vs time}
		\label{fig:temp-depth}
	\end{subfigure}
	\begin{subfigure}[b]{0.19\textwidth}
		\includegraphics[width=\textwidth]{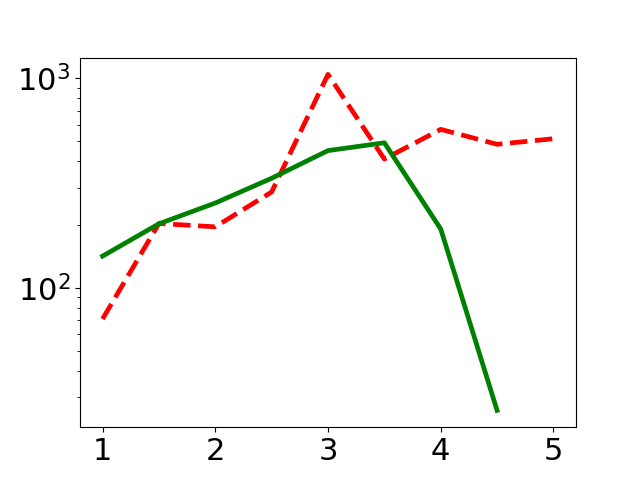}
		\caption{Avg Depth vs time}
		\label{fig:temp-avgdepth}
	\end{subfigure}
	\begin{subfigure}[b]{0.19\textwidth}
		\includegraphics[width=\textwidth]{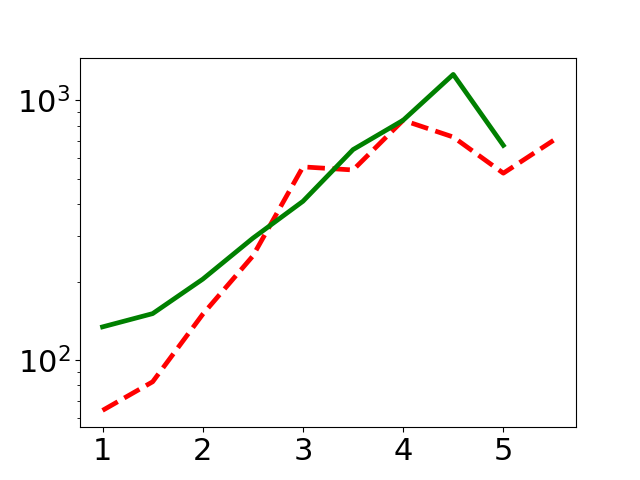}
		\caption{Virality vs time}
		\label{fig:temp-virality}
	\end{subfigure}
\caption{Temporal profiles of diffusion properties of the cascades generated by the posts of the hate and non-hate users. The posts of hateful users spread farther, wider and deeper more quickly in the initial stages.} 
 \label{fig:temporal-diffusion}
\end{figure*}

\noindent\textbf{Temporal evolution of the cascade parameters}: We also explored the different temporal aspects of information diffusion, namely the evolution of different cascade parameters over time. We illustrate the temporal evolution for both KH and NH cascades in terms of size, breadth, depth, average depth and structural virality via the Figures~\ref{fig:temp-size},~\ref{fig:temp-breadth},~\ref{fig:temp-depth},~\ref{fig:temp-avgdepth} and~\ref{fig:temp-virality} respectively. For all such diagrams, the x-axis represents the specific characteristic (like size or depth) and the y-axis represent the time taken in thousand seconds. 
It is quite evident that the time taken for the KH cascades to reach a particular value is lower in the initial stages implying that KH cascades are significantly faster initially. This can be attributed to the high proportion of KH users as early propagators. 

\noindent\textbf{Summary}:
\begin{itemize}
\item The posts of hateful users diffuse significantly farther, wider, deeper and faster than non-hateful ones. 
\item Posts having attachments as well as those exhibiting community aspect tend to be more viral.
\item Hateful users are more proactive and cohesive. This observation is based on their fast repost rate and the high proportion of them being early propagators.
\item Hateful users are also more influential due to the significantly large values of structural virality, average depth and depth. 
\end{itemize}

\section{RELATED WORK}
To the best of our knowledge there has not been any work that tries to study the diffusion of hate in online social media. However, there are several works that looks into diffusion in 
fake news~\cite{mit-media-2018,mustafaraj2017fake,wu2018tracing}, 
Linkedin~\cite{anderson2015global}, 
retweet cascade~\cite{Cheng_cascade_2014,goel2015structural,mit-media-2018,cheng2016cascades}, 
rumours~\cite{friggeri2014rumor,leskovec2007patterns,zhao2015enquiring,jin2013epidemiological,del2016spreading} and 
Tumblr~\cite{alrajebah2017deconstructing,xu2014rolling,chang2014tumblr,alrajebah2015investigating}.~\cite{cheng2016cascades} perform large scale analysis of recurring cascades in Facebook. They observe that content virality is the main driver for recurrence. In ~\cite{del2016spreading}, the authors perform a large scale analysis of Facebook and observe that selective exposure to content is the primary driver of content diffusion and generates the formation echo chambers.

There is little research done on Gab. \cite{zannettou2018gab} performed the first study in which the author collected and analyzed a large dataset of Gab and found that the site is predominantly used for discussion of news, world events, and politics. They also found that Gab contains 2.4 times most hatspeech as compared to Twitter. ~\cite{Lima2018InsideTR} also found that Gab is very politically oriented and users who abuse the lack of moderation disseminate hate.~\cite{zannettou2018origins} perform a large scale measurement study of the meme ecosystem by introducing a novel image processing pipeline. Gab has substantially higher number of posts with racist memes. Gab share hateful and racist memes at a higher rate than manstream communities. Similarly, ~\cite{finkelstein2018quantitative} study millions of comments and images from alt-right web communities like  4chan's Politically Incorrect board (/pol/) and the Twitter clone, Gab and quantify the escalation and spread of antisemitism.



The majority of the research in hatespeech has been done in the aspect of detection in various social media platforms like Twitter~\cite{waseem2016hateful,Qian2018HierarchicalCF,davidsonautomated,Badjatiya:2017:DLH:3041021.3054223}, Facebook ~\cite{del2017hate}, Yahoo! Finance and News~\cite{Warner:2012:DHS:2390374.2390377,Djuric:2015:HSD:2740908.2742760,Nobata:2016:ALD:2872427.2883062}, and Whisper~\cite{mondal2017}. In another online effort, a Canadian NGO, the Sentinel Project\footnote{\url{https://thesentinelproject.org}}, launched a site in 2013 called HateBase~\footref{hatebase}, which invites Internet users to add to a list of slurs and insulting words in many languages. 
There are some works which have tried to characterize the hateful users. In ~\cite{ICWSM1817837}, the authors study the user characteristics of hateful accounts on Twitter and found that the hateful user accounts differ significantly from normal user accounts on the basis of activity, network centrality, and the type of content they produce. In ~\cite{elsherief2018peer}, the authors perform a comparative study of the hatespeech instigators and target users on Twitter. They found that the hate instigators target more popular and high profile Twitter users, which leads to greater online visibility.~\cite{mathew2018thou} looks into the aspects of counterspeech on hateful YouTube video and develops machine learning models to automatically detect counterspeech in YouTube comments. In~\cite{hatelingo2018}, the authors focus on studying the target of the hatespeech - directed and generalized. They observe that while directed hate speech is more personal and directed, informal and express anger, the generalized hate is more of religious type and uses lethal words such as `murder', `exterminate', and `kill'.

\section{Discussion}

\textbf{Account characteristics of hateful and non-hateful users}: We analyze the differences in the account characteristics of hateful and non-hateful users. The different account characteristics include the number of posts, followers and followings (normalized over time) and the number of likes, dislikes, replies, reposts (normalized over the number of posts) of the KH and NH users. The normalization over time is done by dividing the account characteristic (say number of posts) of an user by the number of days elapsed from the first post of the user to the date the last post was crawled. We report the mean and median details of these characteristics in Table~\ref{tab:account-characteristics}. We measure the statistical significance between the two distributions using the two sample K-S test and observe that each of the account characteristics are significantly different ($p$-value\textless0.001).
The inordinate difference in the mean and median values between NH and KH can be attributed to the prolific activity of hateful users. The raw quantity of posts generated by the KH and NH amount to 18.65\% (3.95M) and 52.94\% (11.22M) of all posts respectively. This implies that \textbf{0.3\% of the users generated 18.65\% of the content} and thus KH users are significantly more influential.

\begin{table}[t]
\centering
\resizebox{0.45\textwidth}{!}{%
\begin{tabular}{|c|c|c|c|c|}
\hline
Feature& Mean KH& Mean NH & Median KH& Median NH\\ \hline
post&10.19&0.68&3.49&0.079\\ \hline
follower&1.97&0.67&0.72&0.15\\ \hline
following&1.99&0.95&0.32&0.06\\ \hline

like&2.60&1.54&1.46&0.70\\ \hline
dislike&0.13&0.05&0.04&0.00\\ \hline
score&2.54&1.738&1.41&0.89\\ \hline
reply&0.22&0.12&0.16&0.00\\ \hline
repost&0.35&0.23&0.14&0.00\\ \hline
F:F&4.90&5.05&1.66&1.56\\ \hline

\hline
\end{tabular}}
\caption{Account characteristics of the hateful and the non-hateful users. Hateful users generates more popular content and also posts frequently. All the differences in account characteristics are significant($p$-value\textless0.001).}
\label{tab:account-characteristics}
\end{table}

\noindent\textbf{Network characteristics}: In this section, we look into the network characteristics of the hateful and non-hateful users on the basis of their follower-following relationship. We construct a subgraph over the entire network with nodes being the set of hateful and non-hateful users and edges representing the follower-following relationship between these users only. This subgraph so formed has 63K nodes and 8.33 M edges. We observe that the network of hateful users (1K nodes, 42.4K edges) is $\approx$ 20 times more dense than the non-hateful users (62.8K nodes, 7.45M edges). The hateful users also demonstrate higher reciprocity values (58.3\%) as opposed to the non-hateful users (55.9\%). Moreover, a non-hateful user is 6.8 times more likely to follow a hateful user than a hateful user following a non-hateful user, inkling at the higher popularity of hateful users. It is also 22.75 times more likely that a hateful user will follow another hateful user than a non-hateful one. This indicates strong cohesiveness between the hateful users. 

\noindent\textbf{Case study of Pittsburg shooting}: In the aftermath of the Pittsburg synagoue shooting\footref{pittsburg_shooting}, the owners were forced to shutdown the site temporarily for a week~\footnote{https://www.technadu.com/godaddy-forces-gab-shut-down-temporarily/46040}. The reason behind the decision to ban the website arose from Robert Bowers' history of posting anti-Semitic messages on Gab (under the username @onedingo). Bowers allegedly killed eleven people at a Pittsburgh synagogue with a gun on October 27, 2018. 

We illustrate the account characteristics of @onedingo that were present in our dataset in the Table~\ref{tab: onedingo-details}. We observe that all the characteristics of @onedingo are close to the characteristics of the hateful users shown in Table~\ref{tab:account-characteristics}. We also manually looked into the user's posts and found several hateful instances such as this :``\textit{Kikes are enemy number one. Dealing with anything after will be a relative piece of cake. I will not fire on someone who is shooting my enemy}''. Moreover, not only were 23\% of onedingo's followings but also 14\% of his followers were hateful users. 

\begin{table}[t]
\centering
\begin{tabular}{l l l}
\hline
Account characteristics & Value & Normalized value\\ \hline \hline
post&206&1.355\\ 
follower&212&1.395\\ 
following&232&1.526\\
like&568&2.757\\ 
dislike&2&0.01\\ 
score&566&2.748\\
reply&113&0.549\\
repost&114&0.553\\
F:F&0.91379&- - \\ \hline
\end{tabular}
\caption{Description of onedingo's  account characteristics}
\label{tab: onedingo-details}
\end{table}


\noindent\textbf{Limitations of the current study :} In our analysis we have relied on the user account to study the cascade. We assume that the hateful posts of these hateful accounts would generate majority of the reposts. This means that few of the reposts of these hateful accounts might not be hateful in nature. However, while we cannot claim to have captured the full picture, our analysis provided a peek into the cascade dynamics of the hateful posts in Gab. 




\section{Conclusion and Future work}

In this paper, we perform the first study which looks into the nuances of the diffusion characteristics of the posts made by hateful and non-hateful users. We used high precision keywords to select hateful users and provide them as input to the DeGroot's model to identify the hateful and non-hateful set of users. We then analyse the diffusion characteristics of the posts of these users. We found that the posts made by hateful users tend to spread farther, faster, and wider. These hateful users are densely connected with each other and generate almost 1/5th of the content in Gab despite comprising 0.3\% of the users.

Our work also points to several open research avenues. A large fraction of the posts were in the form of images in case of hate users. For the future work, we could look into the task of building a classification system that can distinguish between images/videos that are hateful in nature. Another interesting direction would be to look into the diffusion characteristics of the individual hateful posts instead of the accounts.

\begin{small}
	\bibliography{Main}

\begin{thebibliography}{}

\bibitem[\protect\citeauthoryear{Alrajebah \bgroup et al\mbox.\egroup
  }{2017}]{alrajebah2017deconstructing}
Alrajebah, N.; Carr, L.; Luczak-Roesch, M.; and Tiropanis, T.
\newblock 2017.
\newblock Deconstructing diffusion on tumblr: structural and temporal aspects.
\newblock In {\em Proceedings of the 2017 ACM on Web Science Conference},
  319--328.
\newblock ACM.

\bibitem[\protect\citeauthoryear{Alrajebah}{2015}]{alrajebah2015investigating}
Alrajebah, N.
\newblock 2015.
\newblock Investigating the structural characteristics of cascades on tumblr.
\newblock In {\em Advances in Social Networks Analysis and Mining (ASONAM),
  2015 IEEE/ACM International Conference on},  910--917.
\newblock IEEE.

\bibitem[\protect\citeauthoryear{Anderson \bgroup et al\mbox.\egroup
  }{2015}]{anderson2015global}
Anderson, A.; Huttenlocher, D.; Kleinberg, J.; Leskovec, J.; and Tiwari, M.
\newblock 2015.
\newblock Global diffusion via cascading invitations: Structure, growth, and
  homophily.
\newblock In {\em Proceedings of the 24th International Conference on World
  Wide Web},  66--76.
\newblock International World Wide Web Conferences Steering Committee.

\bibitem[\protect\citeauthoryear{Badjatiya \bgroup et al\mbox.\egroup
  }{2017}]{Badjatiya:2017:DLH:3041021.3054223}
Badjatiya, P.; Gupta, S.; Gupta, M.; and Varma, V.
\newblock 2017.
\newblock Deep learning for hate speech detection in tweets.
\newblock WWW,  759--760.

\bibitem[\protect\citeauthoryear{Bakshy \bgroup et al\mbox.\egroup
  }{2011}]{bakshy2011everyone}
Bakshy, E.; Hofman, J.~M.; Mason, W.~A.; and Watts, D.~J.
\newblock 2011.
\newblock Everyone's an influencer: quantifying influence on twitter.
\newblock In {\em Proceedings of the fourth ACM international conference on Web
  search and data mining},  65--74.
\newblock ACM.

\bibitem[\protect\citeauthoryear{Chang \bgroup et al\mbox.\egroup
  }{2014}]{chang2014tumblr}
Chang, Y.; Tang, L.; Inagaki, Y.; and Liu, Y.
\newblock 2014.
\newblock What is tumblr: A statistical overview and comparison.
\newblock {\em ACM SIGKDD explorations newsletter} 16(1):21--29.

\bibitem[\protect\citeauthoryear{Cheng \bgroup et al\mbox.\egroup
  }{2014}]{Cheng_cascade_2014}
Cheng, J.; Adamic, L.; Dow, P.~A.; Kleinberg, J.~M.; and Leskovec, J.
\newblock 2014.
\newblock Can cascades be predicted?
\newblock In {\em Proceedings of the 23rd International Conference on World
  Wide Web}, WWW '14,  925--936.
\newblock New York, NY, USA: ACM.

\bibitem[\protect\citeauthoryear{Cheng \bgroup et al\mbox.\egroup
  }{2016}]{cheng2016cascades}
Cheng, J.; Adamic, L.~A.; Kleinberg, J.~M.; and Leskovec, J.
\newblock 2016.
\newblock Do cascades recur?
\newblock In {\em Proceedings of the 25th International Conference on World
  Wide Web},  671--681.
\newblock International World Wide Web Conferences Steering Committee.

\bibitem[\protect\citeauthoryear{Davidson \bgroup et al\mbox.\egroup
  }{2017}]{davidsonautomated}
Davidson, T.; Warmsley, D.; Macy, M.; and Weber, I.
\newblock 2017.
\newblock Automated hate speech detection and the problem of offensive
  language.

\bibitem[\protect\citeauthoryear{de Lima \bgroup et al\mbox.\egroup
  }{2018}]{Lima2018InsideTR}
de~Lima, L. R.~P.; Reis, J. C.~S.; Melo, P.~F.; Murai, F.; Silva, L.~A.;
  Vikatos, P.; and Benevenuto, F.
\newblock 2018.
\newblock Inside the right-leaning echo chambers: Characterizing gab, an
  unmoderated social system.
\newblock {\em ASONAM}.

\bibitem[\protect\citeauthoryear{Del~Vicario \bgroup et al\mbox.\egroup
  }{2016}]{del2016spreading}
Del~Vicario, M.; Bessi, A.; Zollo, F.; Petroni, F.; Scala, A.; Caldarelli, G.;
  Stanley, H.~E.; and Quattrociocchi, W.
\newblock 2016.
\newblock The spreading of misinformation online.
\newblock {\em Proceedings of the National Academy of Sciences}
  113(3):554--559.

\bibitem[\protect\citeauthoryear{Del~Vigna \bgroup et al\mbox.\egroup
  }{2017}]{del2017hate}
Del~Vigna, F.; Cimino, A.; Dell’Orletta, F.; Petrocchi, M.; and Tesconi, M.
\newblock 2017.
\newblock Hate me, hate me not: Hate speech detection on facebook.

\bibitem[\protect\citeauthoryear{Djuric \bgroup et al\mbox.\egroup
  }{2015}]{Djuric:2015:HSD:2740908.2742760}
Djuric, N.; Zhou, J.; Morris, R.; Grbovic, M.; Radosavljevic, V.; and
  Bhamidipati, N.
\newblock 2015.
\newblock Hate speech detection with comment embeddings.
\newblock In {\em WWW '15 Companion},  29--30.

\bibitem[\protect\citeauthoryear{ElSherief \bgroup et al\mbox.\egroup
  }{2018a}]{hatelingo2018}
ElSherief, M.; Kulkarni, V.; Nguyen, D.; Wang, W.~Y.; and Belding, E.
\newblock 2018a.
\newblock Hate lingo: A target-based linguistic analysis of hate speech in
  social media.
\newblock ICWSM '18.

\bibitem[\protect\citeauthoryear{ElSherief \bgroup et al\mbox.\egroup
  }{2018b}]{elsherief2018peer}
ElSherief, M.; Nilizadeh, S.; Nguyen, D.; Vigna, G.; and Belding, E.
\newblock 2018b.
\newblock Peer to peer hate: Hate speech instigators and their targets.

\bibitem[\protect\citeauthoryear{Finkelstein \bgroup et al\mbox.\egroup
  }{2018}]{finkelstein2018quantitative}
Finkelstein, J.; Zannettou, S.; Bradlyn, B.; and Blackburn, J.
\newblock 2018.
\newblock A quantitative approach to understanding online antisemitism.
\newblock {\em arXiv preprint arXiv:1809.01644}.

\bibitem[\protect\citeauthoryear{Friggeri \bgroup et al\mbox.\egroup
  }{2014}]{friggeri2014rumor}
Friggeri, A.; Adamic, L.~A.; Eckles, D.; and Cheng, J.
\newblock 2014.
\newblock Rumor cascades.
\newblock In {\em ICWSM}.

\bibitem[\protect\citeauthoryear{Goel \bgroup et al\mbox.\egroup
  }{2015}]{goel2015structural}
Goel, S.; Anderson, A.; Hofman, J.; and Watts, D.~J.
\newblock 2015.
\newblock The structural virality of online diffusion.
\newblock {\em Management Science} 62(1):180--196.

\bibitem[\protect\citeauthoryear{Golub and Jackson}{2010}]{golub2010naive}
Golub, B., and Jackson, M.~O.
\newblock 2010.
\newblock Naive learning in social networks and the wisdom of crowds.
\newblock {\em American Economic Journal: Microeconomics} 2(1):112--49.

\bibitem[\protect\citeauthoryear{Jin \bgroup et al\mbox.\egroup
  }{2013}]{jin2013epidemiological}
Jin, F.; Dougherty, E.; Saraf, P.; Cao, Y.; and Ramakrishnan, N.
\newblock 2013.
\newblock Epidemiological modeling of news and rumors on twitter.
\newblock In {\em Proceedings of the 7th Workshop on Social Network Mining and
  Analysis}, ~8.
\newblock ACM.

\bibitem[\protect\citeauthoryear{Leskovec \bgroup et al\mbox.\egroup
  }{2007}]{leskovec2007patterns}
Leskovec, J.; McGlohon, M.; Faloutsos, C.; Glance, N.; and Hurst, M.
\newblock 2007.
\newblock Patterns of cascading behavior in large blog graphs.
\newblock In {\em Proceedings of the 2007 SIAM international conference on data
  mining},  551--556.
\newblock SIAM.

\bibitem[\protect\citeauthoryear{Mathew \bgroup et al\mbox.\egroup
  }{2018}]{mathew2018thou}
Mathew, B.; Tharad, H.; Rajgaria, S.; Singhania, P.; Maity, S.~K.; Goyal, P.;
  and Mukherje, A.
\newblock 2018.
\newblock Thou shalt not hate: Countering online hate speech.
\newblock {\em arXiv preprint arXiv:1808.04409}.

\bibitem[\protect\citeauthoryear{Mondal, Silva, and
  Benevenuto}{2017}]{mondal2017}
Mondal, M.; Silva, L.~A.; and Benevenuto, F.
\newblock 2017.
\newblock A measurement study of hate speech in social media.
\newblock In {\em HT}.

\bibitem[\protect\citeauthoryear{Mustafaraj and
  Metaxas}{2017}]{mustafaraj2017fake}
Mustafaraj, E., and Metaxas, P.~T.
\newblock 2017.
\newblock The fake news spreading plague: was it preventable?
\newblock In {\em Proceedings of the 2017 ACM on Web Science Conference},
  235--239.
\newblock ACM.

\bibitem[\protect\citeauthoryear{Nobata \bgroup et al\mbox.\egroup
  }{2016}]{Nobata:2016:ALD:2872427.2883062}
Nobata, C.; Tetreault, J.; Thomas, A.; Mehdad, Y.; and Chang, Y.
\newblock 2016.
\newblock Abusive language detection in online user content.
\newblock In {\em WWW '16},  145--153.

\bibitem[\protect\citeauthoryear{Qian \bgroup et al\mbox.\egroup
  }{2018}]{Qian2018HierarchicalCF}
Qian, J.; ElSherief, M.; Belding-Royer, E.~M.; and Wang, W.~Y.
\newblock 2018.
\newblock Hierarchical cvae for fine-grained hate speech classification.
\newblock {\em EMNLP} abs/1809.00088.

\bibitem[\protect\citeauthoryear{Ribeiro \bgroup et al\mbox.\egroup
  }{2018a}]{ICWSM1817837}
Ribeiro, M.; Calais, P.; Santos, Y.; Almeida, V.; and Jr., W.~M.
\newblock 2018a.
\newblock Characterizing and detecting hateful users on twitter.

\bibitem[\protect\citeauthoryear{Ribeiro \bgroup et al\mbox.\egroup
  }{2018b}]{riberio18}
Ribeiro, M.~H.; Calais, P.~H.; Santos, Y.~A.; Almeida, V. A.~F.; and Jr., W.~M.
\newblock 2018b.
\newblock Characterizing and detecting hateful users on twitter.
\newblock {\em CoRR} abs/1803.08977.

\bibitem[\protect\citeauthoryear{Taxidou and Fischer}{2014}]{LRIF-2014}
Taxidou, I., and Fischer, P.~M.
\newblock 2014.
\newblock Online analysis of information diffusion in twitter.
\newblock In {\em Proceedings of the 23rd International Conference on World
  Wide Web}, WWW '14 Companion,  1313--1318.
\newblock ACM.

\bibitem[\protect\citeauthoryear{Vosoughi, Roy, and
  Aral}{2018}]{mit-media-2018}
Vosoughi, S.; Roy, D.; and Aral, S.
\newblock 2018.
\newblock The spread of true and false news online.
\newblock {\em Science} 359(6380):1146--1151.

\bibitem[\protect\citeauthoryear{Warner and
  Hirschberg}{2012}]{Warner:2012:DHS:2390374.2390377}
Warner, W., and Hirschberg, J.
\newblock 2012.
\newblock Detecting hate speech on the world wide web.
\newblock In {\em Proceedings of the Second Workshop on Language in Social
  Media}, LSM '12,  19--26.

\bibitem[\protect\citeauthoryear{Waseem and Hovy}{2016}]{waseem2016hateful}
Waseem, Z., and Hovy, D.
\newblock 2016.
\newblock Hateful symbols or hateful people? predictive features for hate
  speech detection on twitter.
\newblock In {\em Proceedings of the NAACL student research workshop},  88--93.

\bibitem[\protect\citeauthoryear{Wu and Liu}{2018}]{wu2018tracing}
Wu, L., and Liu, H.
\newblock 2018.
\newblock Tracing fake-news footprints: Characterizing social media messages by
  how they propagate.
\newblock In {\em Proceedings of the Eleventh ACM International Conference on
  Web Search and Data Mining},  637--645.
\newblock ACM.

\bibitem[\protect\citeauthoryear{Xu \bgroup et al\mbox.\egroup
  }{2014}]{xu2014rolling}
Xu, J.; Compton, R.; Lu, T.-C.; and Allen, D.
\newblock 2014.
\newblock Rolling through tumblr: characterizing behavioral patterns of the
  microblogging platform.
\newblock In {\em Proceedings of the 2014 ACM conference on Web science},
  13--22.
\newblock ACM.

\bibitem[\protect\citeauthoryear{Zannettou \bgroup et al\mbox.\egroup
  }{2018a}]{zannettou2018gab}
Zannettou, S.; Bradlyn, B.; De~Cristofaro, E.; Kwak, H.; Sirivianos, M.;
  Stringini, G.; and Blackburn, J.
\newblock 2018a.
\newblock What is gab: A bastion of free speech or an alt-right echo chamber.
\newblock In {\em Companion of the The Web Conference 2018 on The Web
  Conference 2018},  1007--1014.

\bibitem[\protect\citeauthoryear{Zannettou \bgroup et al\mbox.\egroup
  }{2018b}]{zannettou2018origins}
Zannettou, S.; Caulfield, T.; Blackburn, J.; De~Cristofaro, E.; Sirivianos, M.;
  Stringhini, G.; and Suarez-Tangil, G.
\newblock 2018b.
\newblock On the origins of memes by means of fringe web communities.
\newblock {\em arXiv preprint arXiv:1805.12512}.

\bibitem[\protect\citeauthoryear{Zhao, Resnick, and
  Mei}{2015}]{zhao2015enquiring}
Zhao, Z.; Resnick, P.; and Mei, Q.
\newblock 2015.
\newblock Enquiring minds: Early detection of rumors in social media from
  enquiry posts.
\newblock In {\em Proceedings of the 24th International Conference on World
  Wide Web},  1395--1405.
\newblock International World Wide Web Conferences Steering Committee.

\end{thebibliography}
	\bibliographystyle{aaai}
\end{small}

\end{document}